# The effects of competing magnetic interactions on the electronic properties of $CuCrS_2$ and $CuCrSe_2$

Girish C Tewari, T S Tripathi, A K Rastogi[*]

School of Physical Sciences, Jawaharlal Nehru University, New Delhi 110067, India

E-mail: akr0700@mail.jnu.ac.in

## Abstract

We present a detail study of the electrical resistivity, thermoelectric power, magnetic susceptibility $\chi$ and the heat capacity $C_P$ in antiferromagnetic layered compounds $CuCrS_2$ and $CuCrSe_2$ at 2K-300K. $CuCrS_2$ showed sharp cusp in $\chi$ and a lambda-like peak in $C_P$ at $T_N = 40K$ as expected for a 3D- magnetic order, while more metallic $CuCrSe_2$ showed a rounded maximum in $\chi$ and the absence of sharp peak in $C_P$ around 55K, the $C_P$ at low temperature has $T^2$–dependence in it which suggests the absence of the long range order and 2D spin-liquid like excitation in its magnetic phase. We explain the absence of the magnetic order in the selenide compound as resulting from the effective competition of the magnetic interactions from the distant neighbors; the indirect exchange among the intra-layer Cr-atoms increases in more metallic selenide compound which competes with the direct antiferromagnetic interactions between the Cr-atoms of different layers which destroys the long range magnetic order.

1. Introduction

In the layered compounds of Cr-chalcogenides $ACrX_2$; X= S and Se, the triangular layers of Cr in a sandwich of X-Cr-X alternate with a layer of non-magnetic A-atoms ( Na, K , Cu or Ag) to give hexagonal-rhombohedral structures. The weak interlayer coupling results in the relatively low ordering temperature of $T_N$= 20K to 55K, while interactions among the Cr-atoms of the layer are strong. Different neutron scattering studies reveal non-collinear magnetic ordering as a function of the Cr-separation, resulting from the competition of interactions from the distant neighbors in these compounds [1-5].The direct exchange interaction between the Cr-atoms is antiferromagnetic because of the filled $t_{2g}$ orbitals of the $Cr^{3+}$-ions while the indirect ~$90^0$ Cr-X-Cr superexchange interactions is ferromagnetic, in selenide compounds with the increase in Cr-separation the direct interaction weakens [4, 5]. The asymptotic value of the paramagnetic Curie temperature $\theta_{CW}$ -- related to the dominating intra-layer interactions, was found to vary from a large negative value of -300K to a positive value of +250K with the Cr-Cr separation, respectively in $LiCrS_2$ and $KCrSe_2$ [1-3].

Recently some studies has been directed towards understanding the non-collinear magnetic ordering and associated magneto-elastic properties of $Cu(Ag)CrS_2$ in the model of "the geometric frustration" of a Heisenberg antiferromagnet on a triangular lattice ( HAFT) that is applicable in case of corresponding oxides $AMO_2$ ( M=Ti, Cr, Fe , Ni) , notably $CuCrO_2$ [6-8]. However, in case of more covalent sulfide and selenide compounds a similar model of HAFT may not be applicable, since in these cases the non-

Heisenberg interactions such as the indirect exchange interactions-- superexchange or kinetic exchange interactions via different routes dominate, and it is the competition among them that gives rise to the observed non-collinear helical order. In case of $ACrX_2$ ; X= S and Se , the modulation period of the magnetic moments of the Cr-atoms of the layer is found to increase with Cr-Cr separation and in $KCrSe_2$ they are ferromagnetically ordered while remaining anti-parallel in the alternate layers [3-5].

The electronic transport properties of these compounds, especially those containing Cu and Ag, have not been properly studied. Different studies have reported quite contrasting conductivity behavior; with ρ (300K) = 0.025 Ω-cm in crystals [9] to immeasurably high resistivity of pure $CuCrS_2$ and with the insulator to metal transition by V-substitution of Cr-atoms etc [10]. In our recent work we found that, contrary to previously believed insulating nature, the samples of $CuCrS_2$ are clearly non-insulating but have a complex ρ(T) dependence and showing a minimum in the resistivity with temperature. The thermopower S was remarkably large and nearly constant value at high temperatures [11, 12]. We believe this behavior to be the intrinsic properties of $CuCrS_2$ and can be explained in the model similar to the self doped Kondo-insulator. Here the "Kondo-insulator" like excitation gap can be realized due to the hybridization of narrow 3d-band (instead of 4f-bands of rare–earth) with the valence band of p-orbitals of S or Se atoms in their paramagnetic phase [13, 14].

In this paper we present the electronic transport and magnetic properties of $CuCrSe_2$ and compare them with $CuCrS_2$. Our results clearly show a non-insulating nature of these compounds [11, 12]. The specific heat results of $CuCrSe_2$ are presented for the first time. We show that our selenide compounds are more metallic and in it the magnetic ordering process and the magnetic excitations at low temperatures are qualitatively changed as inferred by the large heat capacity in it that shows a $T^2$-dependence.

2. **Preparation and Characterization**

$CuCrS_2$ and $CuCrSe_2$ were prepared from a direct reaction of pure elements in required molecular ratio in a sealed quartz tube. The Initial reaction when carried out at lower temperatures, by a slow heating to $650^0C$, gave significant contamination of cubic spinel compound $CuCr_2X_4$ which is ferromagnetic above room temperature. After thorough grinding of reacted mass, pelletizing under 5 ton pressure, sintering above $900^0C$ in sealed tube over 5-7 days and followed by air quenching completely eliminated the ferromagnetic phase, as could be checked by the magnetization measurements at room temperature[11, 12]. In some cases the single crystal flakes were also obtained at the colder end of the tube. The use of iodine for the vapor transport in case of selenide, however, gave copper deficient $CrSe_X$–crystals. The compounds were characterized by X-ray diffraction on powders and by EDAX attached to the SEM for the chemical analysis.

In figure 1, we have presented the X-ray diffraction pattern of $CuCrSe_2$, together with the Reitveld refined profile in the space group of R3m using GSAS [15]. The structural parameters are found to be as a=b=3.677Å, c=19.38Å, z(Cr)=0.0, z(Cu)=0.142(2), z(SeI)=0.260(1) and z(SeII)=0.740(1). The fitting of our X-ray pattern is quite satisfactory as can be seen from the reasonable value of $R(F^2)$=0.0854 and $\chi^2$=0.26. The refined values are very similar to the values of Bongers et al [1].

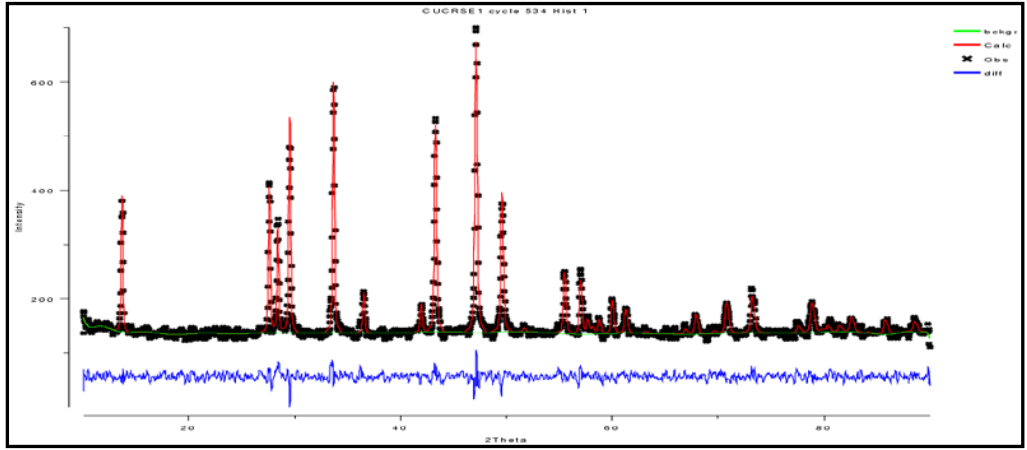

Figure 1: The Reitveld refined profile of X-ray diffraction of $CuCrSe_2$ in the space group of R3m.

### 3. Electronic Transport property

The electrical conductivity measurements were made on thin circular polycrystalline pellets in the van der Pauw geometry and using Ag-paint for the electrical contacts. The Seebeck coefficient S (= $\Delta V/\Delta T$) was measured with respect to copper at closed interval of temperature (15-300K) in an apparatus built in a closed- cycle helium refrigerator system. At each temperature small temperature difference was created across the sample length of 5-7mm. A differential thermocouple (Au (0.05%Fe)/Chromel) gave the temperature difference. The spurious and the offset voltages of the measuring circuit were eliminated by reversing the directions of temperature gradient and averaging the measured thermal voltages.

**3.1** *Resistivity and Thermopower*

The results for $CuCrS_2$ were already reported previously where $\rho$ and S were found to depend on the macro- structural features of the pellets resulting from the preparation conditions; for samples S1(flake), S2 and S3 the $\rho$ and S varied between 5-500m$\Omega$-cm and 150-450$\mu$V/K respectively at room temperature [11, 12]. In an earlier study $CuCrS_2$ and $AgCrS_2$ were also found with contrasting resistivity values of respectively $\rho_{300K}$ = 20$\Omega$-cm and >2x10$^7$ $\Omega$-cm, while they showed same antiferromagnetic transitions respectively at 39K and 42K [8].

In figure 2, we have plotted the electrical resistivity $\rho$ (T) and Seebeck coefficient S (T) of $CuCrSe_2$ and compared our previous results on $CuCrS_2$. The Sample Se1 was twice annealed at 900$^0$C and it does not contain ferromagnetic impurities, which were usually present in as-prepared samples (here Se2). The sample Se3 was quenched from 1100$^0$C-- designated here as H.T. phase. The preliminary results on the magnetic ordering behavior of H. T. sample suggested that the quenching from high temperatures results in the substantial amount of interstitial Cr-atom in the Cu- layer as was also reported for the $CuCrS_2$ [12]. The H.T. quenched phase Se3 showed an increase in the resistivity on cooling.

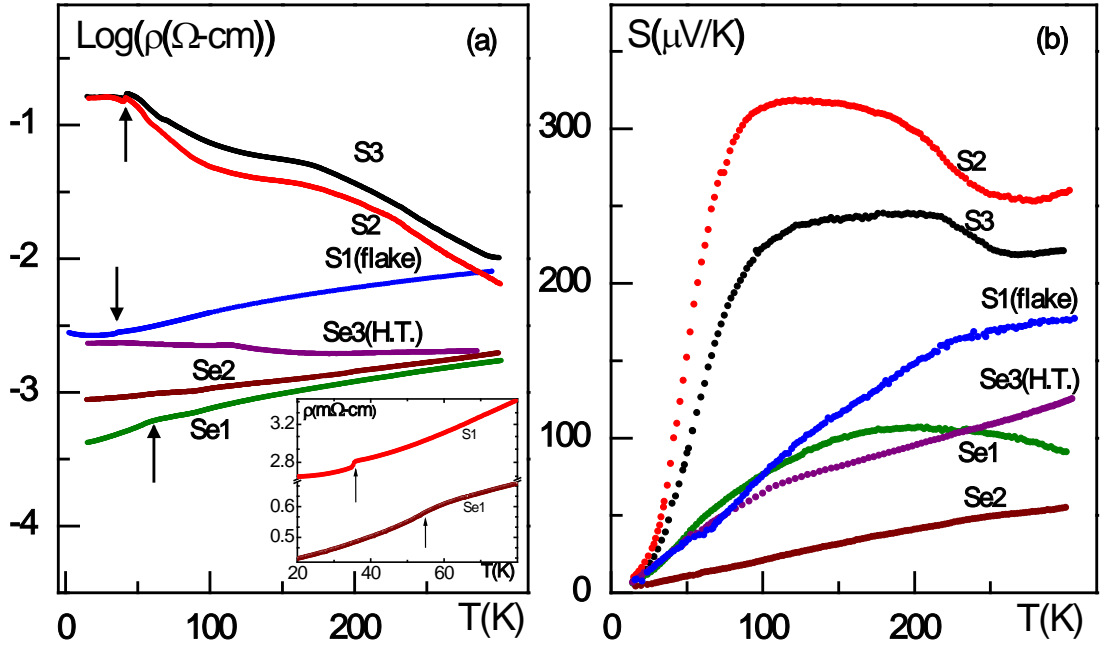

Figure 2: (a) Resistivity ρ and the (b) Seebeck coefficient S of $CuCrS_2$ (S1, S2 and S3-flake) and $CuCrSe_2$ (Se1, Se2 andSe3), in the inset of (a) weak nature of the anomalies in ρ at $T_N$ are shown.

The ρ(T) dependence in figure 2(a) is complex, but the compounds are clearly non-insulating at low temperatures; $CuCrSe_2$ is more conducting with a metallic temperature dependent. In our opinion, the most remarkable transport property of these compounds is that in spite of their metallic nature, the value of S is large and it tends towards a nearly saturated value at high temperature. For more conducting compounds a Kondo-like minimum in ρ(T) is found. We have explained these non-Fermi liquid like features in the transport properties as resulting from the strongly hybridized nature of conduction electron and its scattering from the spin-fluctuations on the 3d-electrons of Cr-atoms, which leads to a Kondo-type energy gap near Fermi energy in their paramagnetic phase [11-14].

### 3.11 *Transition anomalies*

The effect of the magnetic ordering on the resistivity is shown in the inset of figure 2(a). Apart from a small discontinuity in ρ at $T_N$ due to a structural change by the magneto-elastic coupling in case of $CuCrS_2$ [7], the anomalies in ρ and S are relatively weak, hence in these compounds the effects on the conductivity due to the critical scattering or the effects of altered Brillouin - zone by the antiferromagnetic ordering are relatively weak. On the other hand, at $T_N$ =55K a change in dρ/dT can be inferred in case of $CuCrSe_2$, which is surprisingly similar to the resistive anomalies found at the magnetic transition in a metallic ferromagnets like elemental Ni and Gd etc. and also to the chemically related cubic spinels of Cr-chalcogenide -- the $CuCr_2X_4$ (X=S, Se or Te) at their ferromagnetic transition ($T_C$ = 320-460K) [16].

## 4. Magnetic properties

The magnetic properties of $CuCrS_2$ have been variously reported [1-8]. While the ferromagnetic impurities could not permit the earlier study of the magnetic properties of $CuCrSe_2$, the properties of the corresponding $AgCrSe_2$ could be measured [1, 2, 17]. We are able to synthesis a pure phase of $CuCrSe_2$ by the reaction at a higher temperature and after repeated sintering it at $900^0C$. In figure 3(a), we have presented the magnetization results of both the compounds at 1 kOe magnetic field and have also included the results of $AgCrSe_2$ [17], the isothermal magnetization at 1.8K is plotted for both of them in figure 3(b).

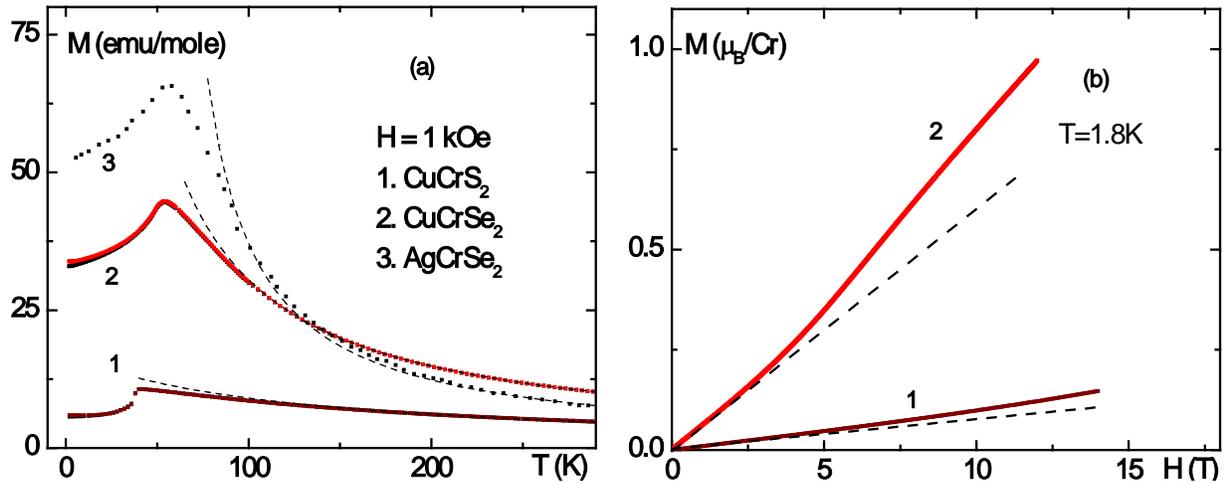

Figure3: (a) The M(T) of $CuCrS_2$ and $CuCrSe_2$ in 1 kOe field showing a reversible maximum under field cooling and zero field cooling. Curie-Weiss dependence (dotted curves) is found above 100K. The data for $AgCrSe_2$ are from the earlier work [17]. (b) The isothermal magnetization at 1.8K showing an upturn above 5T and 1.5T indicating a non-collinear magnetic order.

All compounds, in figure 3(a), show a maximum in the magnetization at low temperatures. However, as compared to the sharp cusp like anomaly at the antiferromagnetic transition of $CuCrS_2$, the maximum in selenide compound is well rounded; the behavior is reversible when measured in 1kOe field after cooling under field (FC) and zero field-cooling (ZFC). The $\chi$ can be fitted to the Curie-Weiss dependence $\chi = C/(T-\theta)$ from 100K to 300K in both the compounds (as marked by the dotted curve) with C equal to 1.9 and 2.9 and $\theta$ equal to negative -110K and positive +5K values respectively for our sulfide and selenide compound. The higher value of C, than the expected value of 1.87 for the $Cr^{3+}$-ions with S=3/2, indicate that the ferromagnetic correlations exist among Cr-atoms in the paramagnetic phase of $CuCrSe_2$ at least up to 300K.This clearly shows that the interactions between the bare-moments of Cr-atoms are much stronger, and a very smaller value of $\theta = +5K$ that was obtained by fitting our data below 300K may be the result of the effective cancellations from the distant neighbors. Earlier, for $AgCrSe_2$ a value of C=1.6 and a larger value of $\theta = +50K$ was deduced from the fitting the data between 316 and 400K [17].

The change from a large negative in sulfides to a positive value of $\theta$ in selenide compounds has been related to the weakening of the direct antiferromagnetic interaction with the increase in Cr-Cr separation, in favor of the indirect ferromagnetic interaction [1, 2]. It is however interesting to note that these large changes have relatively small effects on the ordering temperature; this behavior may be related to the layer structure and the anisotropy of interactions in these compounds. With the competing interactions from the distant neighbors a non-collinear spin configuration is expected, and is clearly indicated by our observation of the upturn in M(H) above the external fields of 5T and 1.5T respectively for $CuCrS_2$ and $CuCrSe_2$ at 1.8K, see figure 3(b).

## 5. Specific Heat

In the earlier studies of the heat capacity $C_P$ of the sulfide compounds $CuCrS_2$ and $AgCrS_2$, a sharp anomaly at their respective magnetic transition $T_N$= 38K and 42K was found [6, 18]. A latent heat contribution of a first order change at its magnetic ordering transition could also be deduced for $AgCrS_2$ and the subsequent neutron and synchrotron diffraction results of $CuCrS_2$ showed it as due to the simultaneous distortion of the lattice to monoclinic symmetry [7]. The change in magnetic entropy in $AgCrS_2$ below $T_N$ was about 70% of the value that is expected for the 3-dimensional magnetic order of the $Cr^{3+}$-spins [18].

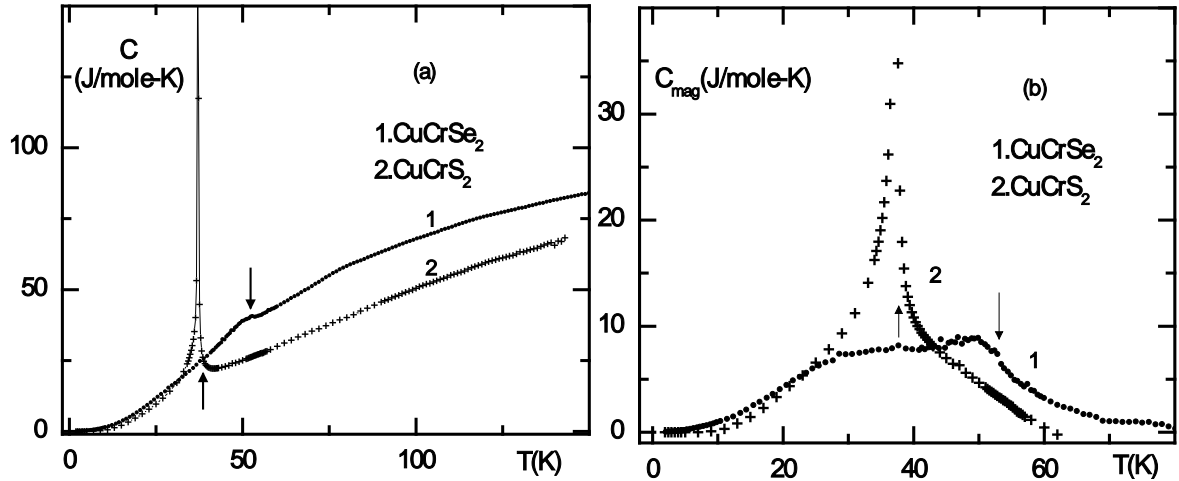

Figure4: (a) Specific heat $C_P(T)$ showing a lambda–like anomaly in $CuCrS_2$ and an extended maximum in $CrCrSe_2$ at magnetic ordering. (b) The large magnetic contribution to heat capacity indicating extensive short range ordering above 55K in $CuCrSe_2$.

In the figure 4(a), we have presented the results of $C_P$ of $CuCrS_2$ and $CuCrSe_2$. In selenide a sharp peak at $T_N$ is absent and is replaced by a visible maximum. In the right panel of figure 4(b), we have plotted the contribution of magnetic heat capacity after subtracting the lattice contribution from the measured $C_P$. The lattice contribution was calculated in the Debye model by using Debye temperature $\theta_D$ = 280 K and 340K respectively for $CuCrSe_2$ and $CuCrS_2$. The arrows in the figure mark the temperature of the susceptibility cusp in $CuCrS_2$ and the rounded maximum in $\chi$ (figure 3(a)) in case of $CuCrSe_2$. In the later the magnetic contribution in $C_P$ extends over to much higher temperature, indicating extensive short range ordering above 55K, and the transition to long range order seems to be blocked at low temperatures probably because of competing interactions causing a short spin-correlation length among the layer atoms.

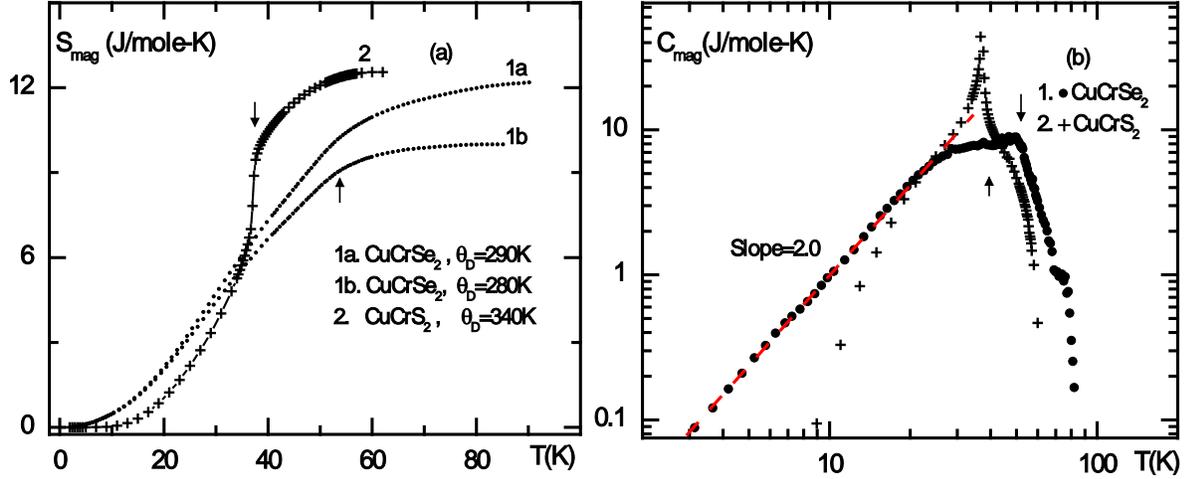

Figure5: (a) Magnetic entropy of CuCrSe$_2$ showing a large value and a continuous increase across 55K. (b) The linear curve of the log-log plots of C$_m$ vs. T for CuCrSe$_2$ at low temperatures with slope 2, giving a T$^2$-dependence of C$_m$ below 14K.

### 5.1 *Magnetic entropy*

The magnetic entropy, S$_{mag}$ is calculated by integrating C$_m$/T versus T curves, and is shown in the left-panel of figure 5 for both the compounds. We have drawn two separate S$_{mag}$(T)-plots for CuCrSe$_2$ after using different values of Debye temperature θ$_D$ = 280K and 290K in order to estimate C$_m$. The magnetic entropy of CuCrS$_2$ is small at low temperatures and increases rapidly near the transition due to the destruction of long range magnetic order at 38K, where it attains about 9J/mole-K — i.e. about 75% of the entropy of the spin disorder at the transition for the localized moments of Cr$^{+3}$-ions with S = 3/2. The neutron diffraction study found a reduced moment of 2.37μ$_B$/Cr-atom in its helical order [5] may account for the entropy of the ordered phase. A visible jump like small increase in S$_{mag}$ at the transition temperature, in figure 5(a), may be associated with the latent heat of the monoclinic distortion which was confirmed by a separate study mentioned above [7].

CuCrSe$_2$ shows a qualitatively different behavior of S$_{mag}$ on heating: S$_{mag}$ is quite large at low temperatures, increases rather slowly on heating and increases continuously across 55K—the temperature of maximum in χ (fig 3(a)), till about 80K. This dependence of S$_{mag}$(T) strongly suggests the absence of long range magnetic order in this compound. We have plotted C$_m$ versus T on a log-log scale in figure 5(b), the plot for CuCrSe$_2$ is linear with a slope of 2, meaning that C$_m$ = AT$^2$ below about T = 14K with coefficient A = 9x10$^{-3}$ J / (mol Cr K$^3$).

### 5.2 $T^2$-dependence

A $T^2$-dependence of the magnetic specific heat with nearly same coefficient as in CuCrSe$_2$ was also reported in CuCr$_{1-x}$V$_x$S$_2$ for x=0.3. In this case, the bond frustration of antiferromagnetic interactions by the atomic substitution results in a spin-glass (SG) like order at a reduced temperature; this is indicated by a characteristic irreversibility in the magnetization on cooling under field [6]. In the selenide compound the irreversibility of a spin glass like disorder is absent (see figure 3(a)). Nevertheless, a $T^2$–dependence in both of them is not natural to a canonical SG phase, where instead a T-dependence below the freezing temperature is universally observed [19]. There is now a vast literature on the elementary excitations of a diluted 2D antiferromagnet (e.g. Kagomé spin glass SrCr$_{9p}$Ga$_{12-9p}$O$_{19}$) wherein the long-range order is inhibited by frustration (geometrical or otherwise) and at low temperatures the collective excitations of the spin-liquid give $C(T) = AT^2$; this dependence was found to be surprisingly robust against the atomic substitution for increasing the bond disorder [20]. The observation of the $T^2$–dependence in both the compounds CuCrSe$_2$ and CuCr$_{0.7}$V$_{0.3}$S$_2$ clearly suggest that the spin-liquid like excitations also exist in case of covalent solids with non-Heisenberg magnetic interactions.

6. **Conclusions**

We have presented, for the first time, the detail electronic properties of CuCrSe$_2$ and have shown that it is more metallic than CuCrS$_2$ and shows similar large thermopower with a saturating behavior at high temperature. The most remarkable features are in its magnetic and specific heat properties. In selenide compound the magnetic ordering process is quite different from the corresponding sulfide; the latter is well known to undergo a 3D long range non-collinear magnetic ordering with accompanying lattice distortion, below 40K. On the other hand, in our pure CuCrSe$_2$ the M(T) and C(T) measurements suggest a complex short range ordering that extends over to the much higher temperature than 55K where M showed maximum. In this case the absence of the irreversibility in M under field cooling clearly discounts spin–glass freezing. Interestingly, here $C_m$ varies as $T^2$ at low temperatures, and surprising similar dependence is also found in CuCr$_{1-x}$V$_x$S$_2$; x=0.3; in the latter the long range non-collinear order is suppressed by the lattice defects and instead a linear T-dependence is expected below the spin glass freezing [6]. A $T^2$–dependence has been discussed in the context of the elementary excitations of the spin-liquid in a diluted antiferromagnet, such as 2D-Kagomé spin glass SrCr$_{9p}$Ga$_{12-9p}$O$_{19}$ [20]. In the light of similar behavior of the largely covalent compounds in which the competition from the long range indirect exchange interaction can also suppress the magnetic order, a deeper understanding of the elementary excitations of the spin- glass (or a spin liquid) state is required.

# 7. Acknowledgment

G.C. Tewari and T. S. Tripathi acknowledge the Council of Scientific and industrial Research, India, for the financial support. The authors acknowledge AIRF, JNU for XRD and SEM (EDAX) measurements. We thank Dr. C. S. Yadav and Prof. P. L. Paulose, TIFR Mumbai for magnetization and specific heat measurements.